\documentclass[onecolumn,aps,preprint,showpacs]{revtex4}%
\usepackage{amsfonts}
\usepackage{amsmath}
\usepackage{amssymb}
\usepackage{graphicx}%
\providecommand{\U}[1]{\protect\rule{.1in}{.1in}}

\begin{document}
\preprint{ }
\title[Target decay on inhomogeneous structures]{Target annihilation by diffusing particles in inhomogeneous geometries}
\author{Davide Cassi}
\affiliation{Dipartimento di Fisica, Universit\`{a} di Parma, viale Usberti 7/a, 43100
Parma, ITALY}
\affiliation{}

\begin{abstract}
The survival probability of immobile targets, annihilated by a population of
random walkers on inhomogeneous discrete structures, such as disordered
solids, glasses, fractals, polymer networks and gels, is analytically
investigated. It is shown that, while it cannot in general be related to the
number of distinct visited points, as in the case of homogeneous lattices, in
the case of bounded coordination numbers its asymptotic behaviour at large
times can still be expressed in terms of the spectral dimension $\widetilde
{d}$, and its exact analytical expression is given. The results show that the
asymptotic survival probability is site independent on recurrent structures
($\widetilde{d}\leq2$), while on transient structures ($\widetilde{d}>2$) it
can strongly depend on the target position, and such a dependence is
explicitly calculated.

\end{abstract}
\volumeyear{2009}
\volumenumber{number}
\issuenumber{number}
\eid{identifier}
\date[Date text]{date}
\received[Received text]{}

\revised[Revised text]{}

\accepted[Accepted text]{}

\published[Published text]{}

\startpage{1}
\endpage{ }
\maketitle

The kinetics of diffusion limited reactions is deeply affected by geometry and
topology, and their description in terms of random walks models is a very
powerful tool to explore such a dependence \cite{weisslibro,avrahav}. In some
cases, especially on regular lattices, it is possible to establish simple
analytical relations between survival probabilities of chemical species and
basic random walks functions, making explicit the dependence on universal
geometrical parameters such as spatial dimension. The case of irregular
structures is more complex, due to the absence of symmetries allowing for a
reduction of the degrees of freedom involved in analytical calculations.
Therefore, in general, it is not possible to relate time decays and survival
probabilities to simple geometrical parameters characterizing the underlying structures.

In this paper, we face a well known kind of reaction, namely the A+B
$\rightarrow$ B process, where the A species is immobile, which is known as
the target reaction (A is called the target) \cite{bzkprb}. This target
reaction model is usually introduced to describe a variety of physical,
chemical and biochemical processes, such as the Williams-Watts dielectric
relaxation in polymers and glasses \cite{pnas1984} as well as the poisoning of
surface catalysts and of immobilized enzymes \cite{smith}. Such a target decay
problem has been extensively studied analytically, since it admits exact
solutions on homogeneous structures, i.e. on structures where all sites are
topologically equivalent \cite{bzkprb,blumenultra}. In these cases, it can be
shown that the target survival probability decays as a negative exponential of
\textit{S(t)}, the mean number of distinct sites visited by a random walk
after time t. The case of inhomogeneous structures has been recently
investigated for some particular networks, namely Small World Networks (SWN)
\cite{jaschblu} and Scale Free Networks (SFN) \cite{galiblu}, showing that
this simple dependence on \textit{S(t) }no longer holds, and that the
deviations from such a behavior are particularly evident at large times. These
networks, besides being inhomogeneous, exhibit a peculiar feature: the
coordination numbers are not bounded from above in the thermodynamic limit: in
other words the maximum coordination number diverges for $N\rightarrow\infty$,
$N$ being the number of sites.

Many interesting real inhomogeneous system, however, don't present such a
property and the maximum coordination number is finite even for $N\rightarrow
\infty:$ this is the case, for example, of disordered solids, glasses,
fractals, polymer networks and gels. All these structures can be
mathematically described in terms of "physical graphs" , and the applications
of ideas and techniques of algebraic graph theory allows us to obtain
analytical results in spite of the lack of invariance and symmetry
\cite{rassegna}. In the following we present the mathematical formulation of
the target decay problem on physical graphs, and we obtain an exact analytical
expression for the target survival probability at large times, showing that it
can be expressed as a negative exponential of $t$ for transient graphs, and of
$t^{\frac{\widetilde{d}}{2}}$ for recurrent graphs, $\widetilde{d}$ being the
graph spectral dimension.

Let us begin with some basic definitions \cite{rassegna}:

A graph $\mathcal{G}$ is a countable set $V$ of vertices (or sites) $(i)$
connected pairwise by a set $E$ of unoriented links (or bonds) $(i,j)=(j,i)$.

The graph topology can be algebraically represented introducing its adjacency
matrix $A_{ij}$ given by:
\begin{equation}
A_{ij}=\left\{
\begin{array}
[c]{cl}%
1 & \mathrm{if}\ (i,j)\in E\cr0 & \mathrm{if}\ (i,j)\not \in E\cr
\end{array}
\right.  \label{defA}%
\end{equation}
and the coordination number of site $i,$ which is the number of nearest
neighbors of $i$, is given by $z_{i}=\sum_{j}A_{ij}$.

The \textit{discrete time simple random walk} on a graph $\mathcal{G}$ is
defined by assuming that at each time step $t$ the walker can only jump to a
nearest neighbor site, and that all nearest neighbor sites can be reached with
the same probability. Therefore, we can define the jumping probabilities
$p_{ij}$ between sites $i$ and $j$ by
\begin{equation}
p_{ij}={\frac{A_{ij}}{z_{i}}}=(Z^{-1}A)_{ij} \label{jumping}%
\end{equation}
where $Z_{ij}=z_{i}\delta_{ij}$.

Now we introduce the functions $P_{ij}(t)$, each representing the probability
of being in site $j$ at time $t$ for a walker starting from site $i$ at time
$0$, and the \textit{first passage probabilities} $F_{ij}(t)$, each
representing, for $j\not =i$, the conditional probability for a walker
starting from $i$ of reaching for the first time the site $j$ in $t$ steps,
and, for $i=j$ , the probability of returning to the starting point $i$ for
the first time after $t$ steps ($F_{ii}(0)=0$).

The fundamental relation between the $P_{ij}(t)$ and the $F_{ij}(t)$ is given
by
\begin{equation}
P_{ij}(t)=\sum_{\tau=0}^{t}F_{ij}(\tau)P_{jj}(t-\tau)+\delta_{ij}\delta
_{t0}.\label{PF}%
\end{equation}

Introducing the generating functions $\tilde{{P}_{ij}}(\lambda)$ and
$\tilde{{F}_{ij}}(\lambda)$ by the definition

\begin{equation}
\tilde{f}(\lambda)=\sum_{t=0}^{\infty}\lambda^{t}f(t) \label{genfun}%
\end{equation}

\bigskip

from eq.(\ref{PF}) we obtain the simpler relation

\bigskip%

\begin{equation}
\tilde{P}_{ij}(\lambda)=\tilde{F}_{ij}(\lambda)\tilde{P}_{jj}(\lambda
)+\delta_{ij} \label{gf2}%
\end{equation}

which will be useful in the proof of our main results.

On infinite graphs, representing real systems in the thermodynamic limit,
$P_{ii}(t)$ vanishes for ${t\rightarrow\infty.}$ \ If the graph can be
embedded in a finite dimensional Euclidean space, and if the coordination
numbers are bounded, i.e. if $\exists\quad z_{max}\ |\ z_{i}\leq
z_{max}\forall i\in V,$ then $P_{ii}(t)$ vanishes tipically as a power law,
whose exponent allows to define the so called (local) \textit{spectral
dimension }$\widetilde{d}$\textit{, }which is the natural generalization of
the Euclidean dimension for dynamical processes \cite{rassegna}:
\begin{equation}
P_{ii}(t)\sim p_{i}t^{-\widetilde{d}/2}\qquad\mathrm{for}\quad{t\rightarrow
\infty}\quad\forall i\in\mathcal{G} \label{dtl}%
\end{equation}

\bigskip with $\widetilde{d}\geq1$, where $p_{i}$ $\equiv p_{0\text{ }}z_{i}$
(i.e. it depends only on the coordination number of $i$) if $\widetilde{d}%
\leq2.$ For $\widetilde{d}\leq2$, $\tilde{{F}_{ij}}(1)=1,$ i.e., the
probability of ever reaching any site starting from any site is 1, and the
graph is called recurrent. For $\widetilde{d}>2$, $\tilde{{F}_{ij}}(1)<1,$and
the graph is called transient.

Now we can define the target decay problem on an infinite graph, following the
formalism introduced in \cite{jaschblu}.

Since each target decays independently of the other ones, we can study the
decay of a single target without loss of generality. Let us suppose a target
molecule A is placed at site $k$, while, at time $t=0$, the B molecules are
randomly and independently distributed over the other sites, with average site
occupation number $q$. The occupation number distribution at each sites turns
out to be Poissonian, and the probability $p(n)$ of finding exactly $n$ B
molecules at a given site is $p(n)=\frac{q^{n}e^{-q}}{n!}.$For $t>0,$ the B
molecules are moving randomly and independently according to the jumping
probabilities (\ref{jumping}), and the target A is annihilated when it is
reached by one of them. Under these hypotheses, it has been shown
\cite{jaschblu} that the survival probability $\Phi_{k}(t)$ of target A at
time $t$ is given by

\bigskip%

\begin{equation}
\Phi_{k}(t)=e^{-q\Theta_{k}(t)}\qquad\label{survival}%
\end{equation}

\bigskip

where

\bigskip%
\begin{equation}
\Theta_{k}(t)=\underset{i\neq k}{%
{\displaystyle\sum}
}\underset{\tau=0}{\overset{t}{%
{\displaystyle\sum}
}}F_{ik}(\tau)\qquad\label{theta}%
\end{equation}

\bigskip

Now, on homogeneous graphs, $F_{ik}(t)=F_{ki}(t)$, and, due to this symmetry,
$\Theta_{k}(t)=S(t)-1$, where $S(t)$ is the number of distinct sites visited
by a walker after $t$ steps (on homogeneous graphs it is independent of the
starting site $k)$. This result gives rise to the well known results obtained
on $d$-dimensional Euclidean lattices, where, for $t\rightarrow\infty$,
$\Theta_{k}(t)\sim\sqrt{t}$ for $d=1$, $\Theta_{k}(t)\sim t/\ln t$ for $d=2,$
and $\Theta_{k}(t)\sim t$ for $d\geq3$ \ \cite{bzkprb}.

On inhomogeneous graphs, $F_{ik}(t)\neq F_{ki}(t)$, and the simple relation
mentioned above no longer holds, giving rise to a more complex behavior
\cite{jaschblu}.

Let us proceed to the calculation of the asymptotic behavior of $\Theta
_{k}(t)$ by introducing its generating function:

\bigskip%

\begin{equation}
\widetilde{\Theta}_{k}(\lambda)=\sum_{t=0}^{\infty}\lambda^{t}\Theta
_{k}(t)=\frac{1}{1-\lambda}\underset{}{\underset{i\neq k}{%
{\displaystyle\sum}
}\tilde{F}_{ik}(\lambda)} \label{gentheta}%
\end{equation}

\bigskip

From (\ref{PF}), for $i\neq k,$ we get \cite{rassegna}%

\begin{equation}
\underset{}{\tilde{F}_{ik}(\lambda)=}\frac{\tilde{P}_{ik}(\lambda)}{\tilde
{P}_{kk}(\lambda)}=\frac{z_{k}}{z_{i}}\frac{\tilde{P}_{ki}(\lambda)}{\tilde
{P}_{kk}(\lambda)} \label{F}%
\end{equation}

\ \ \ \ \ \ \ \ \ \ \ \ \ 

therefore
\begin{equation}
\frac{z_{k}}{z_{\max}}\frac{\tilde{P}_{ki}(\lambda)}{\tilde{P}_{kk}(\lambda
)}\leq\tilde{F}_{ik}(\lambda)\leq\frac{z_{k}}{z_{\min}}\frac{\tilde{P}%
_{ki}(\lambda)}{\tilde{P}_{kk}(\lambda)} \label{ineqF}%
\end{equation}

where $z_{\min}\geq1$ is the minimum coordination number.

Moreover, since $\sum_{i}P_{ki}(t)=1$ for every $t$, we have that $\sum_{i\neq
k}\tilde{P}_{ki}(\lambda)=(1-\lambda)^{-1}-\tilde{P}_{kk}(\lambda).$ Therefore%

\begin{equation}
\frac{z_{k}}{z_{\max}}\left(  \frac{1}{(1-\lambda)^{2}\tilde{P}_{kk}(\lambda
)}-\frac{1}{1-\lambda}\right)  \leq\widetilde{\Theta}_{k}(\lambda)\leq
\frac{z_{k}}{z_{\min}}\left(  \frac{1}{(1-\lambda)^{2}\tilde{P}_{kk}(\lambda
)}-\frac{1}{1-\lambda}\right)  \label{ineqtheta}%
\end{equation}

Now we can proceed to the singularity analysis of $\widetilde{\Theta}%
_{k}(\lambda)$ in order to obtain the asymptotic behavior of $\ \Theta_{k}(t)$
by applying Tauberian theorems \cite{flajolet}.

From (\ref{dtl}), we have

\bigskip%

\begin{equation}
\widetilde{P}_{kk}(\lambda)\underset{\lambda\rightarrow1^{-}}{\rightarrow
}\left\{
\begin{array}
[c]{ll}%
p_{0\text{ }}z_{k}\Gamma\left(  1-\frac{\widetilde{d}}{2}\right)
(1-\lambda)^{\frac{\widetilde{d}}{2}-1} & \text{\ for }\widetilde{d}<2\\
p_{0\text{ }}z_{k}\log(1-\lambda)^{-1} & \text{\ for }\widetilde{d}=2\\
\widetilde{P}_{kk\text{ }}(1) & \text{ for }\widetilde{d}>2
\end{array}
\right.  \label{PKK}%
\end{equation}

\bigskip therefore, from (\ref{ineqtheta}),%

\begin{equation}
\widetilde{\Theta}_{k}(\lambda)\underset{\lambda\rightarrow1^{-}}{\rightarrow
}\left\{
\begin{array}
[c]{ll}%
\frac{1}{\overline{z}_{k}p_{0\text{ }}\Gamma\left(  1-\frac{\widetilde{d}}%
{2}\right)  }\frac{1}{(1-\lambda)^{\frac{\widetilde{d}}{2}+1}} & \text{\ for
}\widetilde{d}<2\\
\frac{1}{\overline{z}_{k}p_{0\text{ }}}_{\text{ }}\frac{1}{(1-\lambda)^{2}%
\log(1-\lambda)^{-1}} & \text{\ for }\widetilde{d}=2\\
\frac{z_{k}}{\overline{z}_{k}\widetilde{P}_{kk\text{ }}(1)}_{\text{ }}\frac
{1}{(1-\lambda)^{2}} & \text{ for }\widetilde{d}>2
\end{array}
\right.  \label{solgen}%
\end{equation}

where $\frac{1}{\overline{z}_{k}}\equiv\underset{\lambda\rightarrow1^{-}}%
{\lim}(1-\lambda)\underset{i}{\sum}\widetilde{P}_{ki}(\lambda)\frac{1}{z_{i}%
}=\underset{\lambda\rightarrow1^{-}}{\lim}\underset{i}{\sum}\widetilde{P}%
_{ki}(\lambda)\frac{1}{z_{i}}/\underset{i}{\sum}\widetilde{P}_{ki}(\lambda)$,
with $z_{\min}$ $\leq\overline{z}_{k}\leq z_{\max}$, is the weighted average
of the inverse coordination numbers $1/z_{i}$, with weights $\widetilde
{P}_{ki}(\lambda)$, for $\lambda\rightarrow1^{-}$. Notice that the hypothesis
of boundedness of $z_{i}$ could be replaced with the weaker condition
$\overline{z}_{k}<\infty,$ leaving our results unchanged. On recurrent graphs,
since $\underset{\lambda\rightarrow1^{-}}{\lim}\frac{\tilde{P}_{ki}(\lambda
)}{\tilde{P}_{hi}(\lambda)}=1$, uniformly in $i$, $\overline{z}_{k}$ turns out
to be site independent: $\overline{z}_{k}=\overline{z}$. The singularities in
$\lambda=1$ in (\ref{solgen}) finally give us the following asymptotic behaviors:

\bigskip%

\begin{equation}
\Theta_{k}(t)\underset{t\rightarrow\infty}{\rightarrow}\left\{
\begin{array}
[c]{ll}%
\frac{\sin\left(  \pi\frac{\widetilde{d}}{2}\right)  }{\overline{z}p_{0\text{
}}\pi\frac{\widetilde{d}}{2}}t^{\frac{\widetilde{d}}{2}} & \text{\ for
}\widetilde{d}<2\\
\frac{1}{\overline{z}p_{0\text{ }}}_{\text{ }}\frac{t}{\log t} & \text{\ for
}\widetilde{d}=2\\
\frac{z_{k}}{\overline{z}_{k}\widetilde{P}_{kk\text{ }}(1)}_{\text{ }}t &
\text{ for }\widetilde{d}>2
\end{array}
\right.  \label{sol}%
\end{equation}

\bigskip

These results deserve some comments. First of all notice that, if the graph is
homogeneous, we exactly recover the usual asymptotic form of $S(t)$: this
happens not only for Euclidean lattices \cite{bzkprb}, but also for regular
ultrametric spaces \cite{blumenultra}, and for Bethe lattices, for which the
spectral dimension turns out to be infinite \cite{bethe}. In other cases, such
as fractal structures, even if $\Theta_{k}(t)$ is different from $S_{k}(t)$,
it turns out to have the same $t$ dependence for $t\rightarrow\infty.$

For completeness sake, we notice that, in some particular cases, it is
possible to have logarithmic corrections to the asymptotic behavior described
by (\ref{dtl}) \cite{fibrati}: these corrections give rise to logarithmic
corrections in (\ref{sol}) too, which are rather simple to calculate; we have
neglected them only for simplicity's sake.

Moreover, for recurrent graphs, the asymptotic behavior of $\Theta_{k}(t)$ is
site independent ($\Theta_{k}(t)\rightarrow\Theta(t)$) even if the structure
is inhomogeneous and the sites are not equivalent. This means that, in spite
of inhomogeneity, the recurrent nature of the structure gives rise to
asymptotic survival probabilities which are independent of the target
position: this is the case, e.g. of many deterministic fractals such as the
Sierpinski gasket and the T-fractal \cite{rassegna}, and also of recurrent
random graphs generated with the constraint of bounded coordination number
\cite{clalu}.

On the other hand, when the graph is transient, $\Theta_{k}(t)$ and $\Phi
_{k}(t)$ are site-dependent even for $t\rightarrow\infty$. Such a dependence,
that only concerns the coefficient of the power of $t$, while the exponent is
the same for all sites, is rather intriguing. Indeed, it is a peculiar feature
of inhomogeneous transient graphs. In fact, only on these structures the
probability $f_{i}$ of ever reaching site $i$, averaged over all possible
starting sites, can depend on $i$. In other words, there can exist sites which
are more likely to be visited than others, even at large times, and such a
property gives rise to the asymptotic site dependence of $\Phi_{i}(t).$

From the point of view of the applications, the site-dependence is quite
interesting, since it means that the target position can affect its survival
probabilities at large times, and, in some cases, it is possible to know which
sites are more likely to survive (notice that $\widetilde{P}_{kk\text{ }}(1)$
is the average number of visits to $k$ for a walker starting from $k$ itself).
It is particularly noteworthy the case of the so called \textit{recurrent on
the average} transient graphs \cite{mwrim}, such as, e.g., NT$_{D}$ ("Nice
Trees of dimension \textit{D}") \cite{ntd}, where $\widetilde{P}_{kk\text{ }%
}(1)$ is always finite but unbounded from above, i.e., for every $r\in%
\mathbb{R}
$, it exists some $k$ such that $\widetilde{P}_{kk\text{ }}(1)>r$. In the case
of the NT$_{D}$, which are trees with branches of unbounded length,
$\widetilde{P}_{kk\text{ }}(1)$ is greater for points lying in longer
branches: therefore a target placed in such sites has a greater asymptotic
survival probability.

The detailed investigation of these aspects is fundamental to design optimal
reaction strategies based on geometry, as well as to understand target decay
processes in complex biological systems \cite{farma}.

\end{document}